\documentclass[12pt]{article}

\usepackage[letterpaper,hmargin=1in,vmargin=1in]{geometry}

\usepackage{graphicx,epstopdf,amsmath,amsfonts,amssymb}

\def\be{\begin{equation}}
\def\ee{\end{equation}}
\def\ba{\begin{eqnarray}}
\def\ea{\end{eqnarray}}
\def\ge{\mathrel{\raise.3ex\hbox{$>$\kern-.75em\lower1ex\hbox{$\sim$}}}}
\def\la{\mathrel{\raise.3ex\hbox{$<$\kern-.75em\lower1ex\hbox{$\sim$}}}}

\def\simgt{\mathrel{\raise.3ex\hbox{$>$\kern-.75em\lower1ex\hbox{$\sim$}}}}
\def\simlt{\mathrel{\raise.3ex\hbox{$<$\kern-.75em\lower1ex\hbox{$\sim$}}}}

\newcommand{\bi}[1]{\bibitem{#1}}
\newcommand{\fr}[2]{\frac{#1}{#2}}

\newcommand{\nc}{\newcommand}

\nc{\gone}{\bar g_{\pi NN}^{(1)}}
\nc{\gzero}{\bar g_{\pi NN}^{(0)}}
\nc{\al}{\alpha}
\nc{\ga}{\gamma}
\nc{\de}{\delta}
\nc{\ep}{\epsilon}
\nc{\ze}{\zeta}
\nc{\et}{\eta}
\nc{\ka}{\kappa}
\nc{\rh}{\rho}
\nc{\si}{\sigma}
\nc{\ta}{\tau}
\nc{\up}{\upsilon}
\nc{\ph}{\phi}
\nc{\ch}{\chi}
\nc{\ps}{\psi}
\nc{\om}{\omega}
\nc{\Ga}{\Gamma}
\nc{\De}{\Delta}
\nc{\La}{\Lambda}
\nc{\Si}{\Sigma}
\nc{\Up}{\Upsilon}
\nc{\Ph}{\Phi}
\nc{\Ps}{\Psi}
\nc{\Om}{\Omega}
\nc{\ptl}{\partial}
\nc{\del}{\nabla}
\nc{\ov}{\overline}
\nc{\newcaption}[1]{\centerline{\parbox{15cm}{\caption{#1}}}}
\nc{\us}{U(1)$_S$}

\def\beq{\begin{equation}}
\def\eeq{\end{equation}}
\def\bmat{\begin{displaymath}}
\def\emat{\end{displaymath}}
\def\bear{\begin{eqnarray}}
\def\eear{\end{eqnarray}}
\def\ba{\begin{eqnarray}}
\def\ea{\end{eqnarray}}
\def\bery{\begin{array}}
\def\ery{\end{array}}
\def\bit{\begin{itemize}}
\def\eit{\end{itemize}}
\def\ben{\begin{enumerate}}
\def\een{\end{enumerate}}
\def\btab{\begin{tabular}}
\def\etab{\end{tabular}}
\def\btbl{\begin{table}}
\def\etbl{\end{table}}
\def\bfig{\begin{figure}[htb]}
\def\efig{\end{figure}}
\def\bpic{\begin{picture}}
\def\epic{\end{picture}}

\def\ga{\mathrel{\raise.3ex\hbox{$>$\kern-.75em\lower1ex\hbox{$\sim$}}}}
\def\la{\mathrel{\raise.3ex\hbox{$<$\kern-.75em\lower1ex\hbox{$\sim$}}}}
\def\gappeq{\mathrel{\rlap {\raise.5ex\hbox{$>$}}
{\lower.5ex\hbox{$\sim$}}}}
\def\lappeq{\mathrel{\rlap{\raise.5ex\hbox{$<$}}
{\lower.5ex\hbox{$\sim$}}}}

\def\gyr{{\rm \, G\kern-0.125em yr}}
\def\mev{{\rm \, Me\kern-0.125em V}}
\def\gev{{\rm \, Ge\kern-0.125em V}}
\def\tev{{\rm \, Te\kern-0.125em V}}

\newcommand{\ud}{\textrm{d}}

\begin{document}

\begin{titlepage}

\setcounter{page}{1}

\vspace*{0.2in}

\begin{center}

\hspace*{-0.6cm}\parbox{17.5cm}{\Large \bf \begin{center}

Solar Gamma Rays Powered by Secluded Dark Matter

\end{center}}

\vspace*{0.5cm}
\normalsize

\vspace*{0.5cm}
\normalsize

{\bf Brian Batell$^{\,(a)}$, Maxim Pospelov$^{\,(a,b)}$, Adam Ritz$^{\,(b)}$, and Yanwen Shang$^{\,(a)}$}

\smallskip
\medskip

$^{\,(a)}${\it Perimeter Institute for Theoretical Physics, Waterloo,
ON, N2J 2W9, Canada}

$^{\,(b)}${\it Department of Physics and Astronomy, University of Victoria, \\
     Victoria, BC, V8P 1A1 Canada}

\smallskip
\end{center}
\vskip0.2in

\centerline{\large\bf Abstract}

Secluded dark matter models, in which WIMPs annihilate first into metastable mediators, 
can present novel indirect detection signatures in the form of gamma rays and fluxes of charged particles 
arriving from directions correlated with the centers of large astrophysical bodies
within the solar system, such as the Sun and larger planets. This naturally occurs 
if the mean free path of the mediator is in excess of the solar (or planetary) radius. We show that
existing constraints from water Cerenkov detectors already provide a novel probe of the 
parameter space of these models, complementary to other sources, with significant scope for 
future improvement from high angular resolution 
gamma-ray telescopes such as Fermi-LAT. Fluxes of charged particles produced in mediator decays 
are also capable of contributing a significant solar system component to the 
spectrum of energetic electrons and positrons, a possibility which can be tested with the directional 
and timing information of PAMELA and Fermi.

\vfil
\leftline{October 2009}
    
\end{titlepage}

\subsection*{1. Introduction}

The search for weakly interacting massive particles (WIMPs) as a 
component of non-baryonic dark matter has become a focal point
of modern particle physics \cite{review}. There are several complementary
experimental and observational approaches to WIMP detection \cite{WIMPS}. 
Direct detection experiments 
probe the terrestrial scattering of WIMPs with nuclei
and typically require low radiation environments to keep backgrounds 
under control. High energy colliders 
such as the Tevatron and the LHC offer the possibility of producing WIMPs and measuring 
their properties in the laboratory, provided 
the challenging missing energy signatures can be disentangled. 
Indirect searches for dark matter annihilating into gamma and cosmic rays in the 
galactic halo are also promising,
although susceptible to various, often uncertain,
astrophysical backgrounds. 
Finally, neutrino telescopes such as Super-Kamiokande and Ice Cube 
can search for indirect evidence of 
the annihilation of WIMPs captured in the core of the Sun and the Earth, in the 
form of an observable muon signature arising from neutrino charged current scattering in the detector. 
While the latter two examples are well-known indirect signatures for any thermal relic WIMP dark matter candidate, more
generic WIMPs forming part of a larger dark sector can lead  to further novel signatures.
In this paper, we demonstrate
that models of secluded dark matter \cite{PRV} present an additional 
observational possibility:
high-energy gamma rays and charged particles arriving from a 
direction tightly correlated with 
the centers of the Sun, Earth and other planets. 
Such novel signatures can be effectively probed
with the powerful new generation of
gamma ray telescopes. 

\begin{figure}
\centering{
\includegraphics[width=0.5\textwidth]{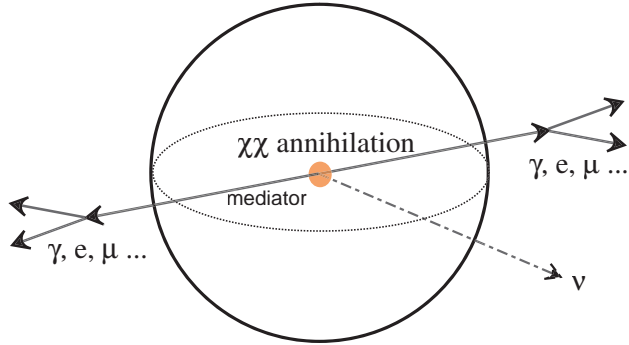}}
\vspace*{-0.0cm}
\caption{\footnotesize A schematic illustration of the new indirect detection signature of secluded WIMPs
captured in the solar core, annihilating to metastable mediators and leading to an electromagnetic flux: $\gamma, e^{\pm},\mu^{\pm},\cdots$. 
Sensitivity to conventional
WIMPs arises only through annihilation to neutrinos.}
\label{fig-scheme}
\end{figure}

The primary feature of secluded models of dark matter \cite{PRV} is a two-stage dark matter annihilation 
process: WIMPs annihilate first into metastable mediators, which subsequently decay 
into Standard Model (SM) states. This breaks the more-or-less rigid link between 
the size of the WIMP annihilation and WIMP-nucleus scattering cross sections. It has been shown that 
a small mass for the mediator allows for new phenomenological possibilities in the form of 
enhanced WIMP annihilation at small velocities \cite{AFSW,PR} that may help to explain various
astrophysical anomalies, e.g. the 
positron excess observed by PAMELA above 10 GeV \cite{pamela}
and perhaps the unexpectedly hard electron spectrum observed by Fermi
above a few hundred GeV \cite{FERMI}. 
Furthermore, a relatively small mediator mass kinematically removes
heavy SM particles from the final state \cite{AFSW,PR}, reconciling these effects with the 
absence of any enhancement in the cosmic ray anti-proton signal \cite{pamela2}.
The lifetime of the mediator is essentially a free 
parameter, limited only by the Big Bang Nucleosynthesis bounds of $\tau \la 1\,$s. 
If this lifetime 
is rather long, the decay of the mediator will occur a long distance away from the 
point of the original WIMP annihilation. Denoting the WIMP particle $\chi$ and the mediator particle $V$,
assuming $\chi\chi\to 2V$ as the main annihilation channel, and taking $m_V \ll m_\chi$,
we arrive at the following estimate for the mediator travel distance:
\be
L = c \tau_V \gamma_V = 3\times 10^6\, {\rm km} \times \fr{ \tau_V }{ 0.01 ~{\rm s}  } \times 
\fr{\gamma_{V}}{10^3}.
\label{L}
\ee
Such large boosts $\gamma_V = m_\chi/m_V$ are easily achieved if the the dark matter mass is near the
electroweak scale and the mediator mass is below a GeV. With regard to the annihilation of WIMPs captured
within the Sun, one can see that 
this distance may very well exceed the solar radius ($R_\odot = 6.96\times 10^{5}$ km)
in which case, unlike conventional WIMPs, most of the decay products 
will not be absorbed. 
For somewhat shorter lifetimes, the interesting possibility emerges of mediators 
produced by WIMP annihilation in the center of the Earth decaying
directly into charged particles within neutrino telescopes. A schematic illustration of this new 
indirect mechanism of probing secluded WIMP models is shown in Fig.~\ref{fig-scheme}. 

A sub-GeV mass mediator will decay predominantly into light states, 
such as pions, muons, electrons, gammas and neutrinos. 
As they are produced in the decays of highly boosted mediators, these decay products 
will be tightly correlated with the direction to the original WIMP annihilation point,
with a typical angular size $\theta_V \sim 1/\gamma_V$. While for charged particles 
this correlation can be reduced by the magnetic fields encountered on the way to the detector, the 
directionality of gammas and neutrinos is unaltered. 
Modern gamma ray telescopes enjoy an angular resolution much better than a degree, which may be exploited 
to enhance the gamma ray signal-to-background. In particular, a notable source of background to 
this signature is the generation 
of gamma rays via cosmic rays impinging on the Sun (see, {\em e.g.} \cite{Seckel,Strong}). Such background gammas 
will typically display a much softer spectrum than gamma rays from  mediator decays and will have no specific 
correlation with the solar center where most of the secluded WIMP annihilation is expected to take place. 
The main gain in sensitivity in detecting gammas from the 
decays of metastable mediators, as compared to a more conventional search for a highly-energetic 
neutrino signal, may come from the increase in efficiency. While the detection of 
multi-GeV neutrinos requires their conversion to muons, which means a loss in efficiency of around ten orders 
of magnitude, the efficiency of detecting gammas created outside of the solar radius 
can be order one.

In addition to gamma rays, many secluded WIMP models with relatively light mediators are destined to produce 
a significant fraction of leptons in the final state, and thus we are naturally led to the question 
of whether mediator decays outside the solar/planetary radii are capable of 
contributing significantly to the fluxes of electrons and positrons seen by PAMELA and Fermi.
It is tempting to pursue the notion that these anomalies may in fact have a local origin
in the solar system, powered by the annihilation of secluded dark matter trapped within solar bodies.
This is an intriguing possibility, as it would offer a new dark matter interpretation of these signatures 
that does not rely on {\it galactic} WIMP annihilation, which generally requires a significant boost factor 
in the annihilation cross section. In this context, a less extreme hierarchy between the WIMP and mediator
mass may be feasible, and 
this idea is akin to using local substructure to enhance the charged particle flux, albeit with a source (the Sun)
which is extremely local and well-understood.  
These electrons and positrons, like the gamma rays discussed above, should be strongly correlated with the center of the Sun, particularly
in the high energy range where the effect of the magnetic fields of the Sun and Earth is less significant. Such a hypothesis appears
straightforwardly falsifiable using directional and timing data from PAMELA and Fermi.
Finally, if charged particles are 
the primary signature, this immediately leads to a minimal gamma ray flux from the associated 3-body decays 
that is also directionally correlated with the center of the Sun, and thus testable by Fermi-LAT.

In this paper we analyze the feasibility of detecting electromagnetic particles, $\gamma$, $e^\pm, \cdots$, arising from the
delayed decays of metastable mediators produced through WIMP annihilation in the deep interior of the Sun and other planetary bodies within the solar system. Our primary focus is on the capability of modern gamma ray telescopes to search 
for these local annihilation signatures and the possibility of distinguishing the signal gammas from 
the solar backgrounds produced by cosmic rays. We demonstrate that in certain corners of the secluded dark matter
parameter space, the gamma-ray signature from the center of the Sun is potentially the most sensitive probe,
superior to direct detection and other indirect signatures. This is especially the case for models 
in which WIMP-nucleus scattering is dominated either by spin-dependent interactions
or proceeds through an inelastic transition to a nearby excited state. We also analyze the 
prospects for detecting gamma rays generated by dark matter annihilations inside Jupiter
using atmospheric \v{C}erenkov gamma-ray detectors, as well as the potential for observing 
pairs of upward going muons in neutrino telescopes generated by the decay of mediators to muons. 
Finally we offer some preliminary speculations concerning an interpretation of PAMELA's rising positron fraction 
originating within the solar system, and discuss the obstacles to 
this identification as well as `smoking gun' directional and temporal signatures which can confirm or 
rule out such an interpretation.

The rest of this paper is organized as follows. In the next section we present the 
WIMP trapping rates and general formulae for calculating the WIMP-powered gamma ray flux, including angular
and energy distributions. Section 3 contains estimates of the gamma ray signature in two variants of secluded 
models, with pseudoscalar(`axion')- and vector-mediation. We conclude in
Section 4 with a discussion of future prospects for improving sensitivity to these dark matter 
models via observation of gamma rays and charged particles in the solar system.

\subsection*{2. Capture and delayed electromagnetic decays of mediators}

In this section we derive the general formulae for the expected gamma ray fluxes generated by 
metastable mediator decays. We will also consider the case of charged particle fluxes and comment 
on the expected effect of the magnetic fields of the Sun and Earth.

\subsubsection*{2.1 Solar capture and $\gamma$-flux} 

We begin by providing the trapping efficiency of WIMPs inside the Sun, 
assuming elastic scattering of WIMPs on nuclei \cite{Gould, Kamionkowski, review}
and normalizing unknown quantities on some fiducial values:
\be
C_{\odot} \simeq 
 1.3 \times 10^{21}~{\rm s}^{-1}  \times  
 \left( \fr{100~{\rm GeV} }{m_{\chi}} \right ) 
\sum_N  f_N  \left(\fr{ \sigma^{\rm SD}_N+ \sigma^{\rm SI}_N}{10^{-42}~{\rm cm}^2}\right) S(m_\chi/m_N) F_N(m_\chi),
\label{scaling}
\ee
where the sum runs over the nuclei $N$ present in the Sun, $f_N$ denotes the fractional abundance relative to hydrogen, $\sigma^{\rm SD}_N$ $(\sigma^{\rm SI}_N)$ is the spin-dependent(spin-independent) scattering cross section and we have used the standard values $\rho_\chi = 0.3$ GeV cm$^{-3}$ and $\bar{v} = 270$ km/s for the
local WIMP density and velocity dispersion.  The function $S(x)=[A(x)^{3/2}/(1+A(x)^{3/2}) ]^{2/3}$, where 
$A(x) = (3/2)[x/(x-1)^2](v_{esc}/{\bar v})^2$, with $v_{esc} \simeq 1156$ km/s an `effective' escape velocity, is a kinematic
suppression factor, while $F_N(m_\chi)$ determines additional suppression from a nuclear form-factor.
This approximate formula holds for both spin-independent and 
spin-dependent scattering, but it is important to bear in mind that
coherent scattering in the former case means that heavier nuclei, such as He, O, and Fe tend to 
dominate the capture rate despite their reduced abundance.  The lack of nuclear 
coherence in spin-dependent scattering means that accounting for
scattering off hydrogen is generally sufficient. More general formulae may be found in Ref. \cite{Gould}.

For most models the trapping rate in the Sun determines the overall annihilation rate, as the 
two processes, trapping and annihilation, are usually in dynamical equilibrium. With an $S$-wave annihilation rate of order 1 pb
as dictated by the relic abundance, a per-nucleon scattering 
cross-section of order $10^{-48}$cm$^2$ or larger is generally sufficient for thermalization to occur over
the lifetime of the Sun. Then the trapping rate $C_{\odot}$ and the probability of a single annihilation event to produce 
gamma quanta reaching the Earth, which we denote as $P_{\gamma}$, 
determines the overall flux of solar WIMP-generated gamma rays at the Earth's 
location:

\begin{eqnarray}
\Phi_{\gamma\odot} = \fr12 \times \fr{C_\odot P_{\gamma} }{4\pi ~ ({\rm A.U.})^2} 
=  1.8 \times 10^{-7} ~{\rm cm}^{-2}~{\rm s}^{-1} \times 
\fr{C_\odot P_{\gamma}}{10^{21}~{\rm s}^{-1}}.
\label{flux}
\end{eqnarray}

This equation is correct if the particles that produce the final photons
are highly boosted along the radial direction of the Sun, and should be
modified by another factor of $1/2$ if those particles travel 
very slowly and emit photons isotropically as the Sun is effectively 
opaque and photons
emitted on the back side of the Sun are reabsorbed. 
We are only interested in the former case in the current discussion.
The master formula (\ref{flux}) suggests that observable gamma 
ray fluxes are indeed possible, provided that $P_\gamma$ is not too small. In conventional
neutralino-like WIMP scenarios, this probability is in fact negligible, as the only practical way of 
producing gamma rays is via high-energy neutrinos interacting with the outer layer of the 
solar material\footnote{Earlier claims of enhanced neutralino annihilation immediately 
outside of the solar radius \cite{early} were not confirmed by subsequent studies \cite{subsequent}.}, and in 
this case the probability $P_\gamma$ is very small. In contrast, secluded WIMPs offer
the  possibility of maximizing $P_\gamma$, which arises
as the product of: (i) the probability $P_{\rm out}$ that the mediator particle decays outside the solar radius;
(ii) the probability ${\rm Br}_{\gamma}$ of producing a gamma quantum in the decay of the mediator; and (iii)
the probability ${\rm Br}_{V}$ of producing the mediator particle in 
the annihilation process:
\be
P_\gamma = g\times P_{\rm out}  \times {\rm Br}_{\gamma} \times {\rm Br}_V.
\label{pgamma}
\ee
Here, $g=g_\gamma g_V$  where $g_\gamma$ and $g_V$ are the multiplicities 
of the photon and the mediator particle $V$ being produced along the chain of
reactions. If the  loss of $V$ due to re-scattering inside the sun can be neglected,
the mediator escape probability $P_{out}$ is well-approximated  by 
\be
P_{out} \simeq \exp\{- R_\odot/(c\tau\gamma)\},
\ee
while the branching ratios ${\rm Br}_{V}$ and ${\rm Br}_{\gamma}$ and the
multiplicities are model dependent. 
For example, ${\rm Br}_{\gamma}$ may be close to 1 in some secluded models, or 
more commonly may lie in the
$10^{-3}-10^{-2}$ range for models where gammas are produced as radiation accompanying 
the decay to charged particles.

The general relations (\ref{scaling}) and (\ref{flux})
use elastic cross sections normalized to $10^{-42} $ cm$^2$, in conflict with the current bounds 
on  spin-independent elastic WIMP-nucleon scattering \cite{SI}, but well below the 
bounds on spin-dependent scattering \cite{SD}. Therefore, the best sensitivity to secluded 
WIMPs will occur for models where nucleon scattering is predominantly spin-dependent.  
Alternatively, scattering may be predominantly  inelastic $\chi_1\to \chi_2$, with 
a small mass gap between the two WIMP states relaxing the most stringent constraints 
on spin-independent $\sigma_p$. We will consider examples in both these categories in Section~3.

\subsubsection*{2.2 Angular and spectral distributions}

We now turn to the angular and spectral distributions of gamma quanta. We will comment at the end of this section on how these distributions may be generalized for charged particles, such as $e^\pm$, for which the propagation effects of the magnetic fields of the Sun and Earth must be taken into account. For simplicity we shall assume a 
decay of the mediator into a pair of gammas, which are monochromatic and isotropic in the rest frame of the mediator. 
The Lorentz boost of the mediator then determines 
both the angular and energy resolution. In the absence of the boost, $\gamma \sim 1$, one would expect the 
Sun to acquire a constant surface brightness in gamma-rays, assuming that the decays happen not far from the 
solar radius, although its side would appear brighter due to simple
geometrical reasons. However, in the most interesting case of large boosts, $\gamma > 10^2$, the gamma quanta 
in the decay products are emitted within an angle $\theta \sim 1/\gamma$ from the original  
direction  of the  mediator. For $\gamma$ as large as $10^3$, the majority of gamma 
rays would come from an angular spot in the sky smaller than the solar radius, and such a tight 
correlation is observable with modern gamma ray telescopes. Also, the gamma energy spectrum 
would be peaked in the direction of the solar center, although this effect is significantly smoothed out
by the finite angular resolution of the detectors.
In this subsection, we illustrate these angular distributions and energy spectra for the case of 
$m_\chi = 1$ TeV WIMPs annihilating into two metastable mediators, that further decay into $2\gamma$, so that the multiplicity factors in Eq. (\ref{pgamma}) are $g_V=g_\gamma=2$ .

In the rest frame of the intermediate particle $V$, photons are emitted 
isotropically with a momentum distribution 
$f'(p')=\frac{2}
{\pi m^2_V}\delta(p'-m_V/2)$, where $p=|\vec p|$.
After a Lorentz transformation, the distribution function $f(\vec p)$ 
measured in the observer's frame becomes
\begin{equation}
f(p,\alpha)=\gamma(1-\beta\cos\alpha)f'[\gamma(1-\beta\cos\alpha)p],
\end{equation}
where $\beta\equiv v/c$, and $\alpha$ is the angle between the 
momentum of the outgoing photon and that of the particle $V$ measured
in the observer's frame. Consequently,
photons are emitted with an angular distribution
given by $1/[4\pi\gamma^2(1-\beta\cos\alpha)^2]$.
The density 
of the photons arriving at the detector with an incoming angle $\theta$
whose momenta are within a small solid angle $\ud\Omega$ is calculated by integrating
the particle density of $V$ along the line of sight weighed by an
appropriate angular distribution factor:
\begin{equation}
\label{eq:better}
\frac{\ud \Phi(\theta)}{\ud \Omega}
=\frac{\cos\theta\;\textrm{Br}_\gamma}{2\pi\tau}\int_D \ud\cos\alpha\;
\frac{n(l\sin\theta\csc\alpha) l \sin\theta}
{\gamma^3\sin^3\alpha(1-\beta \cos\alpha)^2}\,.
\end{equation}
Here $l \approx 1\,\textrm{A.U.}$ is the Sun-detector distance
and $n(r)$ is the number density of $V$ at distance $r$ from the center of the Sun.
Given our assumptions,
\begin{equation}
n(r)=\frac{C_\odot \cdot \textrm{Br}_V}{4\pi v r^2}\exp\{-r/(v\tau\gamma)\}\,,
\end{equation}
where $v\simeq c(1-1/(2\gamma^2))$ is the velocity of $V$.

\begin{figure}[t]
\begin{center}
\includegraphics[width=1\textwidth]
{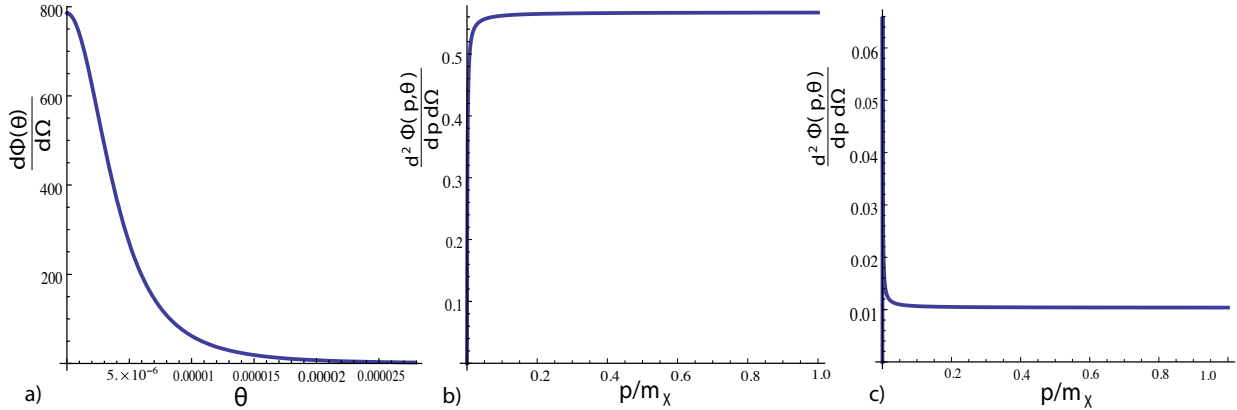}
\caption{\label{fig:spec_smeared}
{\footnotesize Angular and spectral distributions of the photon flux generated by 
two-photon decays of the mediators with $\gamma v\tau=0.5 R_\odot$ and $\gamma=1000$.
(a) The normalized angular distribution $\frac{\ud \Phi(\theta)}{\ud\Omega}/(C_\odot \textrm{Br}_V \textrm{Br}_\gamma
\cdot 10^{-21}\textrm{cm}^{-2})$; and (b) the fully differential flux 
$\frac{\ud^2\Phi(p, \theta)}{\ud p\,\ud\Omega}/
(C_\odot \textrm{Br}_V \textrm{Br}_\gamma
m_\chi^{-1} \cdot 10^{-21}\textrm{cm}^{-2})$ in the direction 
$\theta=0$, averaged over a Gaussian profile that mimics an angular resolution of 
$\Delta\theta=0.1 \theta_\odot$; (c) The same for  $\theta = 0.1 \theta\odot$.}
}
\end{center}
\end{figure}

A few words about the range of the integral for $\ud\cos\alpha$, which
we have denoted as $D$ above, are in order.
If the Sun were transparent, $D$ would simply be 
the domain $[-1,\, +1]$. In the current case, however, the Sun is 
effectively opaque since photons traveling through the interior 
are instantaneously  absorbed or degraded in energy through their interactions with
solar material. Thus one should simply discard those
photons from consideration, which leads to a lower bound
for the integral at
$\cos\alpha=\sqrt{1-(l\sin\theta)^2/R^2_\odot}$ whenever 
$l\sin\theta<R_\odot$. Consequently,
\begin{equation}
D=
\begin{cases}
\left[\sqrt{1-(l\sin\theta)^2/R^2_\odot}\,,\;+1\right],
	& \quad l\sin\theta<R_\odot\;; \\
\left[-1\,,\; +1\right],& \quad l\sin\theta\geq R_\odot\;.
\end{cases}
\end{equation}
The lower integration limits  are not important for
$\gamma\gg 1$, as almost all the photons detected come from within a 
small angle $\sim 1/\gamma$ and the flux is completely negligible 
when $\theta\approx \theta_\odot$.  We illustrate $\ud\Phi/\ud \Omega$
for a representative value of $\gamma=10^3$ in Fig.~\ref{fig:spec_smeared}.
Given a realistic angular resolution, it is clear that all events associated with the 
decays of highly boosted mediators will be concentrated in the angular bin covering the
solar center. 

The energy distribution of $\Phi$ follows from a similar integral. In general,  we can write the full differential
distribution in the form,
\begin{equation}
\begin{split}
\label{eq:spectrum_gen}
\frac{\ud^2 \Phi(\theta, p)}{p^2\ud p \,\ud \Omega}
=&\frac{l\sin\theta \cos\theta \textrm{Br}_\gamma}
{\gamma\tau}\int_D \ud\cos\alpha \cdot
\frac{f(p,\alpha)\, n(l\sin\theta\csc\alpha)}{\sin^3\alpha}\\
=&\frac{l\sin\theta \cos\theta\; \textrm{Br}_\gamma}{\tau}
\int_D \frac{\ud\cos\alpha}{\sin^3\alpha} \cdot
(1-\beta \cos\alpha) f'[\gamma p(1-\beta \cos\alpha)]
n(l\sin\theta\csc\alpha)\,,
\end{split}
\end{equation}
where $D$ is the same domain explained above.
Specializing to the case of two-body decays of $V$ to photons, we have
\begin{equation}
\label{eq:spectrum}
\frac{\ud^2 \Phi(\theta, p)}{p^2\ud p\, \ud\Omega}
=\frac{(\gamma^2-1)\sin\theta\cos\theta l\; \textrm{Br}_\gamma}
{2\pi\gamma\tau(2\gamma p_0 p-p_0^2-p^2)^{3/2}}\cdot\frac{p}{p_0}\cdot
n\left(\frac{\sqrt{\gamma^2-1}\sin\theta l p}
		{\sqrt{2\gamma p_0 p-p_0^2-p^2}}\right)\,,
\end{equation}
where $p_0\equiv m_V/2=m_\chi/(2 \gamma)$ is the photon energy measured
in the rest frame of $V$. When $l\sin\theta\geq R_\odot$,
this expression is valid as long as
\begin{equation}
p\in \left[\;
p_0\sqrt{\frac{1-\beta}{1+\beta}}\;,\;
p_0\sqrt{\frac{1+\beta}{1-\beta}}
\;\right]\;.
\end{equation}
and is understood to vanish for $p$ outside this range. For $l\sin\theta < R_\odot$,
the lower limit is again modified to $p_0 \gamma^{-1}(1-\beta\sqrt{1-l^2\sin^2\theta/R^2_\odot}\,)^{-1}$
due to absorption. 

The resulting photon spectrum varies dramatically within a tiny range of $\theta$ that
for the Lorentz boosts considered here is typically well below the angular sensitivity of any 
gamma ray telescope. 
For a detector with angular resolution poorer than the angular size of the Sun,
the integration of \eqref{eq:spectrum} over $\ud\Omega$ produces 
a flat spectrum as long as $p_0/\gamma\le p\le p_0\sqrt{\frac{1+\beta}{1-\beta}}$.
However, some detectors, such as Fermi, have an angular resolution
smaller than the solar angular size. For demonstration purposes, we take
$\gamma\sim 1000$, and average \eqref{eq:spectrum} assuming a
Gaussian-profile and an angular resolution $\Delta \theta$ 
of  one-tenth the solar size. Since
$\gamma$ is quite large, almost all the photons that reach the detector come from
within 
an angle of $1/\gamma$ that is comparable to the smallest angular size the detector can resolve.
Consequently, the signal that would be seen by the detector represents a (relatively) bright 
central spot of the Sun with an almost exactly flat spectrum. 
The photon flux drops rather quickly with angle. 
If detectable, the photon spectrum at $\theta > \Delta \theta$ is again mostly flat except for
a very sharp peak near the low momentum end. We illustrate these results in Fig. \ref{fig:spec_smeared}.

Up to this point we have only discussed the spectrum of photons resulting from the decays of 
long-lived mediators. A similar analysis for charged particles, such as electrons and positrons, is less 
straightforward due to the complications of the magnetic fields of the Sun and Earth, as well as the solar wind, and 
their effects on the propagation, energy degradation, and absorption. A proper calculation of the angular and energy 
distributions is beyond the scope of this work, but we wish to give a qualitative discussion of these effects, paying 
particular attention to the implications regarding a possible local solar system component to the PAMELA signal.

A variety of secluded models may be constructed which can lead to electrons and positrons in the final state.
It is important to stress that the effects of the magnetic fields and solar atmosphere on these charged states is 
quite model dependent, and primarily sensitive to the lifetime and boost of the mediator produced in the 
WIMP annihilation as well as the production modes of the final state charged particles. 
If the mediators have a typical decay length on the order of the solar radius or slightly less, processes 
such as absorption and reflection have the potential to strongly 
degrade the overall signal flux, and the strong solar magnetic fields may drastically alter the trajectories of the charged 
particles. If the mediator escapes and decays well past the solar radius, the effects of the heliosphere and 
the Earth's magnetosphere can be still be significant, especially for less energetic particles in the tens of GeV range and below 
(where the bulk of the anomalous PAMELA positrons reside). 

We must emphasize that the main feature of this signal 
would be a high degree of anisotropy in the predicted flux of electrons and positrons, correlated 
with the center of the sun, which should make this scenario testable. 
An important consideration is to what degree this tight correlation may be 
affected or degraded through the processes discussed above. 
It is instructive to compare this signal with a possible pulsar component to 
the rising positron fraction of PAMELA \cite{pulsar}, where one of the main signatures may  again
be an anisotropic signal. While the latter anisotropy may be somewhat difficult to detect,
this feature in the solar component discussed here would be more pronounced due to the
proximity of the Sun and the underlying production mechanism. This provides additional
motivation for  studies of the positional, directional, and temporal features of the existing PAMELA and Fermi data sets.  

In order to go further and actually attempt to fit the PAMELA and Fermi data, a better understanding of the predicted 
energy spectrum would be required, which goes beyond the scope of this paper. Of course, 
this is also quite model dependent. For example, a simple underlying two-body decay of the mediator to $e^+e^-$ pairs 
will result in a very hard spectrum, which may not fit the observed spectrum. However, the spectrum can be softened 
through cascade decays, or the re-scattering of mediators inside the solar interior. It would also be important to understand 
to what extent the electron/positron energy spectrum is affected or softened by interaction with the solar atmosphere and 
through propagation from the sun to the earth.

\subsubsection*{2.3 Other local sources}

The Sun represents by far the largest astronomical body in the solar system, and presumably 
the most efficient WIMP capturing reservoir. However, the lifetime of the mediators can be such that 
their decays occur deep inside the solar interior and the decays outside are exponentially suppressed.  
Capture by planets may then be important and two 
cases to consider are WIMPs captured by the Earth and Jupiter. 
The latter option is of interest because of the possibility of very precise observations
by means of atmospheric \v{C}erenkov detectors. 
The flux of WIMP-generated mediators can be estimated using the basic scaling of (\ref{scaling}) 
suitably rescaling the escape velocity, the mass, and the distance to the Earth. Using 
the results of \cite{Gould}, and assuming for a moment that $V$ decays happen outside the solar interior,
we find
\be
\fr{\Phi_J }{\Phi_\odot} \sim \left(\fr{v_J}{v_{\odot}} \right)^4\times \fr{M_J}{M_\odot} \times 
\left(\fr{L_{S-E}}{L_{J-E}}\right)^2 \sim 4 \times 10^{-9},
\ee
which represents a strong suppression. However, if the decay length of the mediator is less 
than $5\%$ of the solar radius, the gamma ray flux from Jupiter may exceed that from the Sun. 
In this case the use of atmospheric \v{C}erenkov detectors such as HESS, MAGIC and VERITAS, could 
set additional bounds on the capture rate by Jupiter, and consequently on the 
WIMP-nucleus cross section.

Finally, the fluxes from the center of the Earth are difficult to present in the same compact 
form as (\ref{scaling}), primarily because of the likely dependence on the 
annihilation rate. These fluxes are typically much smaller than those generated by WIMP
annihilation in the solar interior  except for some special cases with resonant energy loss
\cite{Gould}. Nevertheless, the annihilation of secluded 
WIMPs inside the Earth does allow a probe of much shorter mediator lifetimes,
and offers additional signatures in neutrino detectors. In certain models, the signature involves
pair production of muons via the decays of the mediators, a possibility that can be efficiently 
explored with neutrino telescopes such as SuperK and IceCube.

\subsubsection*{2.4 Observational sensitivity}

Some of the primary $\gamma$-ray observatories, such as atmospheric \v{C}erenkov detectors
are of limited utility in this case, as they are unable to directly observe the Sun. While, as noted above,
they can be used to place limits on the flux  from Jupiter, a fiducial WIMP-nucleon 
cross section of $10^{-40}$cm$^2$ would generate a  flux that 
is generally too low to provide competitive sensitivity from this source. Thus, we will focus on water \v{C}erenkov detectors and 
space-borne $\gamma$-ray observatories such as Fermi.

\begin{figure}
\centering{
\vspace*{-1cm}
\includegraphics[width=0.55\textwidth]{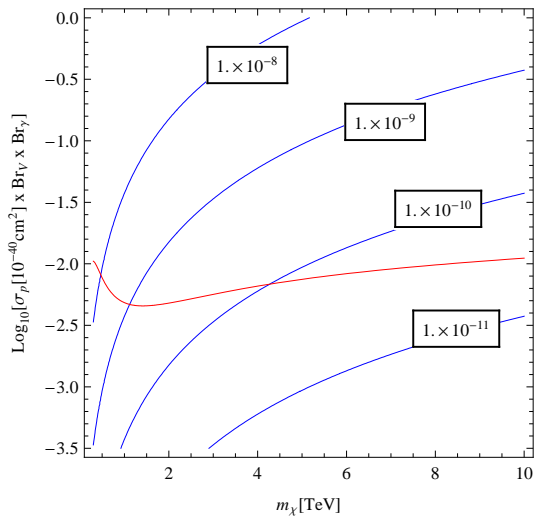}}
\vspace*{-1cm}
\caption{\footnotesize Contours (in blue) of the local $\gamma$-ray flux $\Ph$ in units of cm$^{2}$s$^{-1}$ 
in the plane of dark matter mass $m_\chi$
and a normalized injection cross-section assuming spin-dependent scattering. The decay distance in the 
detector frame has been fixed to $R_\odot$. The red line
indicates a figure of merit for the sensitivity of Milagro (see the text for more details).}
\label{fig-milagro_lim}
\end{figure}

The primary limit we will use here comes from the Milagro water \v{C}erenkov detector \cite{milagro}, which
has a wide field of view and an angular resolution of 0.75$^\circ$, slightly larger than
the disc of the Sun. The background arises primarily from the scattering of high-energy
cosmic rays,  producing $\gamma$'s which can arrive at the detector from within the Sun's
disc. Nonetheless, the limits obtained by Milagro for monochromatic sources are significant
and reach up to 10 TeV.
We exhibit the ensuing constraint on a contour plot of the 
$\gamma$-flux in Fig.~\ref{fig-milagro_lim}, which for characteristic scattering cross-sections,
already imposes a significant constraint on the electromagnetic branching fractions. The primary flux limits
obtained by Milagro are for monochromatic sources, while a monochromatic decay of the mediator
as discussed above leads to a smoothed spectrum.  Given that the 
background is a steeply-falling function of energy, while 
the expected signal is not, we adopt the Milagro bounds 
on monochromatic sources, weakening it by an order of magnitude 
in obtaining the bound in Fig.~\ref{fig-milagro_lim}.

In the near future, the Fermi-LAT detector, which is also able to observe the Sun \cite{lat}
with an angular resolution about 10 times better than that of
Milagro, should be able to improve on these limits and thus provide a very significant probe of
secluded dark matter models. The energy range of Fermi-LAT is also ideally suited to probing 
secluded WIMP annihilation with masses of   a few hundred GeV, where Milagro loses its sensitivity. 

It is also possible for long-lived mediators to decay predominantly to electrons and positrons. 
It is tempting then to consider the implications of these decays in relation to the anomalous positrons fraction
observed by PAMELA. As far as we are aware detailed timing and directional
analyses have not yet been performed, leaving open the possibility of an intriguing local explanation for
these anomalies within the dark matter framework.
Given that the existing magnetic fields affect propagation, and the decay chains of the mediators are 
model-dependent, one could in principle fit the 
spectral shapes of the observed signals. We have obtained a rough estimate of the required integrated 
signal flux by using Fermi electron data with an $\sim E^{-3}$ spectrum to infer an estimate of the 
background and incorporating a new source with a harder power-law spectrum. Depending on the source as well as where 
the signal turns on (i.e. how large a component of the PAMELA signal) we find a required flux of 
\be
\Phi_{e^+e^-}(E>10 ~ {\rm GeV}) \sim 10^{-4} - 10^{-6} ~{\rm cm}^{-2} {\rm s}^{-1}
\label{interesting}
\ee   
We also point out that there will be an ${\cal O}(1)$ reduction of the observed signal flux due to the fraction of the 
time the satellite is facing away from the sun. Given the assumption that at least part of the excess positron flux does arise 
from mediator decays, this leads to a  {\em minimal} predicted gamma flux due to the accompanying final state radiation. 
The typical spectrum of photons resulting from this process would be $(\alpha/\pi )dE/E$, and thus the photon 
flux will generally be no less than 0.1\% of the electron and positron flux, and one infers the following target photon flux:
\be
\Phi^{\rm min}_\gamma(E>10 ~ {\rm GeV}) \ga 10^{-7} - 10^{-9} ~{\rm cm}^{-2} {\rm s}^{-1}.
\label{interesting2}
\ee
Thus, while it is possible that the hypothesis of a local origin for the flux anomalies can be directly tested
with timing and directional information, it also appears that existing EGRET data \cite{egret, Strong} 
may already probe the larger range of this associated gamma flux, while it is feasible that Fermi-LAT will  
be able to cover the entire range of possible fluxes, 
with a potential $10^{-9} ~{\rm cm}^{-2} {\rm s}^{-1} $ level 
sensitivity to multi-GeV gamma 
rays originating from point sources within a year.

\subsection*{3. Secluded models vs $\gamma$-rays}

Having discussed the available sensitivity to $\gamma$-rays from the Sun, in this section we will consider
some secluded dark matter scenarios which would be subject to this indirect probe. 
Given that a large variety of model-building possibilities for WIMPs and mediators have been 
shown to exist \cite{PRV}, we will simply exhibit two classes of models for which the $\gamma$-ray flux
due to WIMP trapping in the Sun may, for various parameters, be the most sensitive observable and 
source of constraints. Our approach will be to fix the parameters such that the decay length of the 
mediator is sufficient to escape the Sun, and then consider
whether the scattering cross-section leading to capture results in a measurable 
(or constrained) $\gamma$-ray flux according to the limits discussed in the preceding section.

\subsubsection*{3.1 Secluded WIMPs with pseudoscalar mediation}

A fermionic dark matter candidate $\chi$ can be secluded by mediating its interaction
with the SM via a pseudoscalar `axion' field $a$ with $m_a<m_\ch$. 
We will imagine a light WIMP, $m_\chi \sim 10 \,{\rm GeV}$, as well as a very light mediator,
$m_a <$ 10 MeV. The interactions comprise a series of dimension 5 operators:
\begin{eqnarray}
{\cal L} &=& {\cal L}_{\rm SM} + \bar \chi (i\partial_\mu\gamma_\mu - m_\chi) \chi +\fr12 (\partial_\mu a)^2 -\fr12 m_a^2 a^2
\nonumber \\
&+& \partial_\mu a \left( \fr{1}{f_\chi} \bar \chi \gamma_\mu \gamma_5 \chi 
+ \sum_q \fr{1}{f_q}\bar q \gamma_\mu \gamma_5 q  + \sum_l \fr{1}{f_l} \bar l \gamma_\mu \gamma_5 l  \right) +\frac{\al}{4\pi f_\gamma} a F_{\mu\nu}\tilde{F}^{\mu\nu}.
\label{axion}
\end{eqnarray}
The coupling constants $f_i$ are above the electroweak scale 
but otherwise are completely arbitrary at this point.
Since this is an effective field theory model, depending on the 
actual UV completion one can achieve the suppression of either 
$f_l^{-1}$ or $f_q^{-1}$ or both (see, {\em e.g.} \cite{PRV2}). The axion can be very long-lived if its
mass is low enough to ensure that decays to hadrons and heavier leptons are kinematically forbidden. 

We will now briefly discuss the interaction rates pertinent to the $\gamma$-ray signal. 
We note that axion mediation with the WIMP sector has also been explored recently in Ref. \cite{Nomura}.

\begin{itemize}
 \item {\it Annihilation}
 
 Unless $f_\ch$ is parametrically larger than $f_f$, $\bar{\chi}{\chi} \rightarrow a a$ will be the dominant annihilation mode given $m_a \ll m_\chi$,
 and the thermally averaged rate is 
 \be
 \langle \sigma v \rangle = \frac{\beta^2}{12 \pi } \frac{m_\chi^2}{f_\chi^4} 
  \rightarrow 2.4 \times 10^{ -26} {\rm cm}^3{\rm s}^{-1},
\label{sigaa}
\ee
where $\beta = (1-4 m_\chi^2/s)^{1/2}$ is the velocity of the WIMPs in the c.o.m. frame. Note that the 
annihilation is in the $P-$wave, a consequence of identical bosons with overall even parity in the final state.
At freeze-out,  taking the WIMP velocity to be $\beta^2 \approx 3 T_f^2/m_\chi^2 \approx 3/20$, the 
relic density requires $f_\chi \simeq 120\,{\rm GeV} \times (m_\chi/10\, {\rm GeV})^{1/2}$.

\item {\it Pseudoscalar Decays}

For axions in the MeV range, we can consider decays to photons and electrons,  
\be
 \Gamma_{a\rightarrow \gamma\gamma} = \frac{ \alpha^2 m_a^3 }{64 \pi^3 f_\gamma^2 }, \;\;\;\;\;\;\; \Gamma_{a\rightarrow e^+ e^-} = \frac{ m_a m_e^2}{2\pi f_e^2}.
 \ee
Depending on the ratios $f_e/f_\gamma$ and $m_a/m_e$, either the photon or electron branching may dominate the 
total width. To maximize the photon fraction, we shall assume that $f_\gamma \sim 10$ TeV, while 
$f_e > {\rm few} \times 10^3$ TeV, in which case the decays are dominated by 
the 2-photon final states. Moreover, provided the characteristic coupling of the axions to quarks is small, $f_q > 100$ TeV, the axions will be long-lived with a 
decay length sufficient to escape the Sun, 
\begin{equation}
L_a=  c\tau_a \gamma_a  \simeq 1.2 \times  10^7 \, {\rm km} \times \left( \frac{m_\chi}{ 10\,{\rm GeV} } \right)
\left( \frac{5\,{\rm MeV}}{m_a} \right)^4 \left( \frac{f_\gamma}{10\,{\rm TeV}} \right)^2,
\label{travel}
\end{equation}
where $\gamma_a \simeq m_\chi/ m_a \simeq 2\times 10^3$ for a fiducial normalization of masses and couplings. 
For the same normalization, one can explicitly check that the absorption cross section of energetic axions in the solar medium is too small to attenuate the flux.

\item {\it Scattering}

The axion-like pseudoscalar mediates spin-dependent WIMP-nucleus elastic scattering, involving the 
effective axion-nucleon couplings  ${\cal L}_{a(n,p)} = (\tilde{f}_p^{-1} \bar{p} \gamma^\mu \gamma^5 p + \tilde{f}^{-1}_n \bar{n} \gamma^\mu \gamma^5 n)\ptl_\mu a$
where $1/\tilde{f}_{(n,p)} = \sum_q \De^{(n,p)}_q/f_q$ in terms of the parameters measuring the spin-content of the nucleons,
 $\Delta_u^{(p)}= \Delta_d^{(n)}\simeq 0.8$, $\Delta_d^{(p)}= \Delta_u^{(n)}\simeq -0.5$, $\Delta_s^{(p)}= \Delta_s^{(n)} \simeq -0.15$. A straightforward calculation
 leads to the tree-level cross-section, conventionally re-expressed in terms of the `model independent' WIMP-nucleon cross section, 
\begin{eqnarray}
\sigma_{p,n} & \equiv & \frac{ \mu^2_{p,n} }{\pi f_\ch^2 \tilde{f}_{(p,n)}^2} \times 
\begin{cases}
1  \qquad\qquad\quad {\rm for} \qquad m_a^2\ll 4\mu_N^2v^2, \\
\displaystyle{\frac{16 \mu_N^4 v^4}{3 m_a^4}} \qquad {\rm for} \qquad m_a^2 \gg 4\mu_N^2 v^2,
\end{cases}
\end{eqnarray}
where $\mu_N$ is the reduced mass for the WIMP-nucleus system.  Note that depending on the mass of 
the mediator, WIMP, and type of nucleus involved in the scattering, there is 
either an enhancement or a suppression. For the trapping rate, we need only consider WIMP scattering with hydrogen 
as the scattering is spin-dependent, 
and in the specific case of the Sun, we must also
account for the increase of the characteristic c.o.m. 
velocity (by a factor of 3-5) due to the WIMPs falling into the Sun's gravitational well. 
The characteristic momentum transfer in this case is
$|{\bf q}| \sim 2 m_N v \approx 8$ MeV, leading us to consider very light mediators in 
the MeV range to avoid a possible velocity suppression.
A light pseudoscalar with mass in the few MeV range will mediate an enhanced interaction due to the long range force. 
Effectively the momentum dependence cancels and we have the usual form for the spin-dependent cross section: 
\begin{equation}
\sigma_p \simeq 2.5 \times 10^{-44}\,{\rm  cm}^2 \times \left( \frac{500\,{\rm  TeV}}{\widetilde{f}_p} \right)^2 
\left( \frac{10 \,{\rm GeV}}{m_\chi} \right),
\label{axcross}
\end{equation}
where we have used the relic abundance to relate $f_\ch$ to $m_\ch$, and 
we have normalized the effective nucleon couplings to be consistent with the 
stringent constraints arising from rare $K$ decays, as we will discuss below.
We see that direct detection constraints are easily satisfied, as Eq. (\ref{axcross}) displays a 
spin-dependent WIMP-nucleon cross section far smaller than the $\sigma_{(p,n)}  < 10^{-37}$ cm$^2$ limit \cite{SD}.

\end{itemize}

With these results in hand, we see that for light WIMPs and pseudoscalar mediators the solar $\gamma$-ray flux 
can be appreciable. For such light WIMPs in the GeV range, we see from Eq.~(\ref{scaling}) 
that the trapping rate can be enhanced by several orders of magnitude compared to weak scale WIMPs, but this is compensated by 
a generically smaller-spin dependent cross section shown in Eq. (\ref{axcross}).

Interestingly such light mediators are not in conflict with astrophysical bounds when the scale
of the interactions $f_q$ is below about $10^6$ GeV, because these states will then thermalize in the core of supernovae
and so will not be subject to the stringent constraints from cooling. Constraints from BBN cannot rule out one additional
thermalized degree of freedom, but in any case a mass in the few MeV range is sufficient to avoid these constraints
entirely. Finally, as alluded to above, rare Kaon decays, in particular searches for 
$K^+ \rightarrow \pi^+ a$, constrain $f_q$ to be above 
100 TeV \cite{Kaon,PRV2,Nomura}, but pose no particular 
problems for the estimate of the spin-dependent scattering 
cross section in Eq.~(\ref{axcross}).

Using Eqs. (\ref{scaling}), (\ref{flux}), and (\ref{axcross}), and assuming an ${\cal O}(1)$ branching of 
axions to photons and a decay length of order the solar radius as discussed above, 
we obtain the characteristic gamma ray flux in the secluded model with axion mediation: 
\begin{equation}
\Phi_{\gamma\odot} \sim 6 \times 10^{-8}\, {\rm cm}^{-2}\, {\rm s}^{-1} \times \left( \frac{10 \, {\rm GeV}}{m_\chi} \right)^2 
\left( \frac{500 {\rm TeV}}{ \widetilde{f}_p }\right)^2.
\end{equation}
This is a large flux which is already close to the range probed by satellites like EGRET (although the mass scale is below the
sensitivity range of Milagro). We have focused on 
light WIMPs and mediators, but different parameter ranges may also yield appreciable fluxes at the expense of 
some fine-tuning of couplings in the quark sector. In particular, a
suppression of the coupling to the top quark, with larger light quark couplings, will relax 
the prohibitive Kaon decay constraints and allow 
an enhanced capture rate for larger WIMP and mediator masses.

\subsubsection*{3.2 Secluded WIMPs with vector mediation}

Unlike the axion-mediated models, secluded WIMPs lying in a hidden sector with
a spontaneously broken \us\ gauge symmetry do  not require additional UV completion, and this sector
naturally couples to the Standard Model through the kinetic mixing portal. WIMP scenarios in this framework are 
straightforwardly formulated \cite{PRV}, and have been a focal point of theoretical interest in the last year due
to the positron data released by PAMELA \cite{pamela}.
A general class of WIMP models involve multi-component states $\ch$, charged under the \us\ vector mediator, and
the low energy Lagrangian after symmetry breaking involving the WIMP, vector $V$, and the Higgs $h'$, takes the form,
\ba
\label{u1}
{\cal L} &=& {\cal L}_{\rm SM} + \bar \chi (iD_\mu\gamma_\mu - m_\chi) \chi + {\cal L}_{\De m}
-\fr14 V_{\mu \nu}^2 +\frac{1}{2}m_V^2 V_\mu^2 + \kappa V_\mu \partial_\nu F_{\mu\nu}  \nonumber\\
&+&  \fr12 (\partial_\mu h')^2 - \fr12 m_{h'}^2( h')^2 +\frac{m_V^2}{v'} h' V_\mu^2 +\cdots
\ea
If charge-conjugation symmetry is broken via ${\cal L}_{\De m}$, the Majorana components of the WIMP
may be split in mass by $\De m \sim \lambda m_V/e'$, which can reduce the elastic scattering cross-section
and ameliorate constraints on $\ka$ from direct detection. The remaining particle physics constraints 
require that $\ka$ be below a ${\rm few} \times 10^{-3}$. 

The lifetime and decay channels for $V$ and $h'$ were analyzed in \cite{BPR}. There are two regimes 
in which either $V$ or $h'$ can be very long-lived. The first 
refers to $m_{h'} > m_V$ and $\kappa < 10^{-9}$. In this case the Higgs$'$ is short-lived, while 
the vector may have lifetimes in excess of a millisecond. This case is of no interest for us in this 
paper, because the trapping rate will scale as $\kappa^2$
 and will be extremely small. The second case with long-lived particles is $m_{h'} < m_V$ and $\kappa \ga 10^{-3}$.
 In this case the extreme longevity of $h'$ comes from the fact that its decay may only 
 proceed at second order in the mixing angle $\kappa$. This kinematic relation renders $h'$ extremely
 long-lived even for moderately small $\kappa$, while the scattering cross section and hence the 
trapping rate can remain large. The relevant interaction rates in this case are detailed below.

\begin{itemize}

\item {\it Annihilation}

Once trapped and accumulated in the center of the Sun, WIMP annihilation may lead to various
final states: $VV$, $Vh'$, and $VVV$. The latter may only be possible 
when annihilation proceeds via capture into an $S=1$ WIMP-onium state \cite{PR},
but otherwise this annihilation cross section is suppressed by an extra coupling
constant. Comparison of the $VV$ and $Vh'$ final state branching is straightforward, 
once we fix the charge assignment for the Higgs$'$ particle and the spin for the WIMP. 
If we assume unit charge under \us\ for the 
complex Higgs$'$ field and a fermionic WIMP, then a comparison of the two final states gives:
\be
\fr{1}{4} \leq \fr{\langle \sigma_{\chi\chi \to Vh'} ~v \rangle }{\langle \sigma_{\chi\chi \to VV} ~ v\rangle } 
\leq 3,
\ee
where the brackets include an average over the spin orientation. 
The upper end of this ratio is achieved  when the annihilation proceeds via the 
formation of WIMP-onium in which case the final state with total spin 1, decaying to $Vh'$ 
is three times more likely than total spin 0, that decays to $2V$. The lower end corresponds 
to the case when the recombination into  WIMP-onium is kinematically forbidden. 
Notice that this ratio is not changed 
by  Coulomb (Sommerfeld) enhancement of the cross section, as it is identical in both channels. 
Thus, in this model a minimum of 
one per every 5 annihilation events results in the production of a possibly very long-lived 
$h'$ particle that is boosted by $m_\chi/m_{h'}$. Note that there is no strict constraint from ensuring
the correct relic abundance in this case, as we can take this as a relation which fixes the 
U(1)$'$ coupling $\al'$, leaving the WIMP mass as a free parameter. For the remainder of this section we will assume 
the $\chi\chi \rightarrow VV$ mode dominates, in which case 
$\alpha'  \sim 0.02\times (m_\chi/500\, {\rm  GeV})$ \cite{PRV}.

\item {\it Higgs$'$ decays}

As noted above, the regime of interest here is when $m_{h'}< m_V$ and the Higgs$'$ is long-lived,
with the dominant decay $h' \to l\bar{l}$ occurring at order $\Gamma_h \sim \kappa^4 \times ({\rm loop~factor})^2$ \cite{BPR}. 
Given $m_V \gg  m_{h'} \gg 2 m_f$ and a boost  $\gamma_{h'} \simeq m_\ch/m_{h'}$,  the $h'$ decay length is 
\begin{equation}
L_h = c\tau_{h'}\gamma_{h'} \sim  10^7\, {\rm km } \times 
\begin{cases}
 \displaystyle{ \left( \frac{\kappa}{5\times 10^{-4}} \right)^{-4} \left( \frac{ m_{h'} }{ 500\,{\rm MeV}} \right)^{-2} 
\left( \frac{ m_{V} }{ 5\,{\rm GeV}} \right)^2  \,
 \qquad {\rm for} \quad m_{h'}>  2 m_\mu}\\
\\
\displaystyle{ \left( \frac{\kappa}{ 5\times 10^{-3}} \right)^{-4} \left( \frac{ m_{h'} }{ 100\,{\rm MeV}} \right)^{-2} 
\left( \frac{ m_{V} }{ 500\,{\rm MeV}} \right)^2 
\, \quad {\rm for}  \quad m_{h'} <  2 m_\mu},
\end{cases}
\label{twocases}
\end{equation}
where $\al'$ has been chosen to fix the relic abundance, 
and we have assumed a WIMP mass of 500 GeV. However, the longevity of the Higgs$'$ boson 
does not guarantee its safe passage through the interior of the sun  as there is a 
potential loss mechanism due to `inverse-Primakoff'
type conversion into $V$ on nuclei, $h'+N\to V+N$, followed by prompt $V$ decay \cite{BPR3}. The cross section for 
this process on protons can be estimated as follows,
\be
\sigma_{abs} \sim 2\pi\alpha\alpha'\kappa^2 m_V^{-2} \sim 4\times 10^{-38} {\rm cm}^2 \times
\left( \frac{\kappa}{5\times 10^{-4}} \right)^2 
\left( \frac{1 \,{\rm GeV}}{ m_V }\right)^2.
\label{absorption}
\ee
This ${ \cal O}(10^{-38}~{\rm cm}^2)$ scale for the cross section 
is sufficiently small that absorption of Higgs$'$ will be negligible, but an increase by two orders of magnitude 
would indeed lead to a significant loss of $h'$ in the solar interior. In particular, a choice of parameters 
as in the second line of (\ref{twocases}) will result in attenuation of the $h'$ flux by more than an order of magnitude. 
Nonetheless, we see that over much of the parameter space $L_h$ can naturally be large enough for $h'$'s to 
escape the Sun without scattering even for relatively large values of $\ka$.

Since the dominant decays are electromagnetic, 
there will be significant photon production through various processes such as internal bremsstrahlung etc. 
However, it is of particular  interest to know the branching to
$\gamma$'s originating from sequential decays of pairs of neutral pions in the product of one-loop induced 
$h'$ decay, or through
a two-loop decay directly to $\gamma\gamma$. The direct decay to $\gamma\gamma$ will have the hardest spectrum, 
while the $\pi_0$-mediated decay is likely to have larger photonic yield than internal 
bremsstrahlung.
We can estimate these branchings by constructing an effective Lagrangian as follows.
We first integrate out all quarks that are heavier than the vector. This leads to an effective interaction of Euler-Heisenberg type  
$\sim {\rm loop} \times (\kappa^2 e^4 /m_Q^4) (F_{\mu\nu})^2 (V_{\alpha\beta}^2)$, 
and similarly a coupling between vectors and gluons. The contribution of such heavy quarks to the relevant 
decays will thus be suppressed by powers of $(m_V/m_Q)^4$ which we will neglect. We are left with the 
light quarks $u,d,s$ as well as perhaps $c, b$ if they are lighter than the vector. 
The next step is to integrate out the vector from the theory at the loop level, which leads to the effective Lagrangian:
\begin{equation}
 {\cal L}_{\psi} = -  c_{ \psi} \sum_{f} Q_f^2 \frac{m_f}{v'} h'  \overline{\psi}_f  \psi_f,
\label{psi}
\end{equation}
where $c_{\psi} = - 3 \kappa^2 \alpha/2 \pi$, and the sum runs over all quarks lighter than the vector.
From here we can straightforwardly compute the decay  $h'\rightarrow \gamma\gamma$ much as in the SM, with the result
\begin{equation}
 \Gamma_{h'\rightarrow \gamma\gamma} = \frac{\alpha' \alpha^4 \kappa^4}{64\pi^4}\frac{m_{h'}^3}{m_V^2} 
\left( \sum_{f} \bigg\vert N_c Q_f^4 I\left( \frac{m_{h'}^2}{  4 m_f^2} \right)\bigg\vert^2  \right),
\end{equation}
where the sum is over all fermions lighter than the vector, and $I$ is the familiar form factor arising from the 
triangle diagram. The factor in the parentheses is of order one. For a heavy Higgs$'$, $m_{h'}> 2 m_\mu$ the 
dominant decay mode is $h'\rightarrow \mu^+\mu^-$, and the branching to a pair of photons is quite small, on the order of $10^{-4}$.
However if $h'$ is lighter than the two muon threshold, $m_{h'}< 2 m_\mu$ the dominant decay mode is $h'\rightarrow e^+ e^-$ with a 
smaller total width. The branching to two photon pairs in this case can be sizable:
\begin{equation}
 {\rm Br}_{h' \rightarrow \gamma\gamma } \simeq   10^{-2} \times \left( \frac{m_{h'}}{100\;\rm MeV}\right)^2 \qquad {\rm for} \qquad
m_{h'} < 2 m_\mu. 
\end{equation}

For a heavy $h'$ it is sill possible to get a large source of gammas through intermediate decays $h'\rightarrow 2 \pi^0$, followed
by the $\pi^0$ fragmenting to photons. The calculation parallels the decay of a light SM Higgs to pions, as  detailed in 
Refs. \cite{voloshin, lighthiggs}. We consider the theory at even lower energies, below the charm mass, 
where we can write the effective Lagrangian in terms of the trace of the QCD energy momentum tensor and 
match on to a chiral Lagrangian. It is then straightforward to calculate the partial width for $h'\rightarrow 2 \pi^0$:
\begin{equation}
 \Gamma_{h'\rightarrow \pi^0 \pi^0}= \frac{\alpha' \alpha^2 \kappa^4 }{2^3 3^4 \pi^3 } \frac{m_{h'}^3}{m_V^2} 
\left( 1- \frac{4 m_\pi^2}{m_{h'}^2} \right)^{1/2} |G(m_{h'})|^2, 
\end{equation}
where the function $G$ is
\begin{equation}
 G(m_{h'}) \equiv
\left\{ 
\left( \sum_F Q_F^2  \right) 
+ \frac{m_\pi^2}{m_{h'}^2} 
\left[ \left( \sum_F Q_F^2  \right)+\frac{3}{2} \frac{4z+1}{1+z}  \right]     
\right\},
\end{equation}
with $z \equiv m_u /m_d \sim 0.56$ and the sum is over the charm and bottom quarks if they are lighter than the vector.
This function is numerically ${\cal O}(1)$ for the parameters of interest here. The $2\pi^0$ partial width is somewhat smaller 
than the $h'\rightarrow \mu\mu$ mode, with a branching of
\begin{equation}
{\rm Br}_{h' \rightarrow \pi^0 \pi^0 } \simeq  5 \times 10^{-2} \times \left( \frac{m_{h'}}{500\;\rm MeV}\right)^2 .
\end{equation}
Thus approximately five percent of all Higgs$'$ decays will result in 4 photons when $m_{h'}$ is somewhat larger than $2 m_{\pi}$. We conclude that both for a light or heavy $h'$ it is possible to obtain percent level branchings into photons.

\item {\it Scattering}

The scattering of WIMPs with nuclei in the secluded vector model was discussed in detail in \cite{BPR2}, and the regime
of most interest here is when a small splitting $\De m$ allows for 1st-order inelastic scattering
on heavier elements. The interesting feature here is that the larger gravitational potential well of the Sun
boosts the c.o.m. kinetic energy of 
the WIMP-nucleus system  relative to the case for terrestrial scattering. This leads to a window for $\De m$ which 
maximizes the trapping rate in the Sun, but which is outside the kinematic
range for terrestrial direct detection. Given $E_{\rm kin} \mu_N,~\Delta m \mu_N \ll m_V^2$:
\ba
\sigma_{\rm inel} = \fr{16\pi Z^2 \alpha\alpha' \kappa^2 \mu_N^2}{m_V^4}
\sqrt{1- \fr{\Delta m}{E_{\rm kin}}},
\ea
which is a simple modification of scattering induced by a finite charge radius for a Dirac WIMP. However,
this requires that $\De m < E_{\rm kin} \sim m_\ch^2 v^2/(2\mu_N)$. The trapping of inelastically scattering
WIMPs, first studied in \cite{inel} was recently considered in detail in \cite{inel1,inel2}, and we can directly make use of their results in
the present case. Indeed, the trapping in the Sun dominantly occurs through scattering on Fe nuclei,
which are lighter than Ge used for example in CDMS. However, the larger velocity relevant for trapping
in the Sun, means that $E_{\rm kin}$ can be larger allowing for a window in $\De m$ unconstrained
by the direct detection limit. Indeed, the cross-section above can be large,
\be
 \sigma_{\rm p}^{({\rm Fe})} \sim 1\times 10^{-39} \, {\rm cm}^2 \times 
\left( \frac{\kappa}{5 \times 10^{ -4 }}\right)^2 
\left(\frac{{\rm 5~GeV}}{m_V}\right)^4 \left( \frac{m_\chi}{500 \, {\rm GeV}}\right),
\ee
where we have exhibited the equivalent per-nucleon cross-section for scattering off Fe nuclei, with $\al'$ traded for the WIMP mass through the relic abundance constraint and $m_\ch \gg m_N$.
Direct detection constraints then arise from 2nd-order 
elastic scattering which for large mediator masses in the GeV range are relatively mild, restricting
$\ka$ to be below 0.1 \cite{BPR2}. 

An alternative route to maximize the trapping rate while satisfying the direct detection 
constraints would be to  introduce 
several (two or more) sequentially mixed $U(1)$ groups, all broken at a sub-GeV scale. This 
would ensure that the WIMP-nucleon cross section contains higher powers of $q^2$ and the resulting form factor 
suppresses coherent scattering on nuclei \cite{formfactor}. At the same time, due to the increased velocity
inside the Sun, the degree of suppression in the capture rate will be much smaller. 

\end{itemize}

Given these results, the  observable $\gamma$-ray flux then follows from the 
$h'\rightarrow \gamma\gamma$ branching fraction ${\rm Br}_\gamma \sim 10^{-2}$ discussed above and the 
dominant contribution from the WIMP-iron inelastic scattering cross section $\sigma^{(Fe)} \sim 10^{-33}\,{\rm cm}^2$ 
to the trapping rate. This easily leads to a detectable flux of gammas for a wide range of parameters,
\begin{equation}
 \Phi_{\gamma\odot} \sim  1 \times 10^{-6}\, {\rm cm}^{-2}\, {\rm s}^{-1} \times \left(\frac{\ka}{5\times 10^{-4}}\right)^2
\left( \frac{5 \,{ \rm GeV}}{ m_V }\right)^4.
\label{result}
\end{equation}
We see that this model leads to a  remarkably large gamma ray flux, that seemingly would 
have been observed by EGRET and Fermi-LAT. Moreover, for the range $m_{h'}> 2 m_\mu$, the accompanying muon flux will generate 
a flux of $\nu_\mu$ well in excess of the bounds set by {\em e.g.} Super-Kamiokande \cite{SuperK}.  
However, it is clear that generic choices of parameters may equally well reduce the flux (\ref{result})
by several orders of magnitude. A more comprehensive  scan of the parameter space for this model vs the resulting 
flux is beyond the scope of this paper, but it clear that the gamma flux does indeed constitute a sensitive probe.

\subsection*{4. Discussion}

Our exploratory study in this paper has suggested a number of striking indirect signatures associated with the annihilation
of secluded dark matter trapped in the Sun (and planets), in the form of a high energy flux of electromagnetic particles 
tightly correlated with the solar center. In this section, we will finish with a number of additional remarks.

One of the issues that has become apparent is that there are relatively few competitive limits on  gamma rays from the Sun, primarily because
this is a difficult source to handle for many of the more sophisticated gamma-ray telescopes.  However, 
Fermi-LAT can and has observed the Sun and its impressive angular resolution and long exposure times 
point to it as having the best experimental sensitivity for probing the gamma-ray signatures of  
secluded dark matter scenarios discussed here, with mediator lifetimes on the scale of the solar radius. 
 Indeed, according to our analysis of model scenarios in Section~3, Fermi-LAT may well 
 provide sensitivity to specific parts of the parameter space which is superior to all other direct or indirect probes.

Beyond direct decays to photons,  secluded models also naturally produce a significant branching of mediators to high-energy electron-positron pairs,
As discussed earlier, its an intriguing possibility that massive bodies in the Solar System may provide another possible source 
of the anomalous electron and positron fluxes in the multi-GeV range. Such scenarios could be probed via high-precision analyses of the 
spatial or temporal non-uniformities in these fluxes, particularly the positron fraction observed by PAMELA. 
Studies of this kind could significantly strengthen the parameter reach in 
many models of secluded dark matter. 
Moreover, while the requisite flux of charged particles 
could be achieved in many variants of secluded dark matter, they most likely will be put to the test by the 
upcoming Fermi gamma-ray data  given the minimal flux that arises through final state radiation.

Within this general framework, we have presented two concrete models of secluded dark matter consistent with 
the relic abundance and direct detection constraints in which the most accessible experimental signatures are gamma rays 
and charged particles from the Sun. The crucial ingredient is a mediator with a lifetime long enough to escape 
from the Sun, but short enough on  cosmological or astrophysical scales so that
there exist no additional constraints arising from early cosmology beyond those already 
considered in the literature (see, {\em e.g.} \cite{cmb}). It is likely that other models of this type may be constructed, 
and indeed  it would be worthwhile to explore a more generic scan to understand if the 
gamma-ray flux observed in these specific models is robust.

Beyond these points that were already touched upon in the text, we would also like to mention some other 
related issues concerning indirect detection signatures:

\begin{itemize}

\item Secluded mediators with long lifetimes can, depending on the particular model and mass range under consideration, 
greatly enhance the overall neutrino flux reaching the Earth. For example, if the mediators are heavy enough to decay 
to charged mesons outside the
solar radius, the production of muon neutrinos is inevitable. Indeed, even if this decay occurs inside
the solar radius, the neutrino yield can be significant provided  the decays occur outside the 
dense core where all muons and pions are quickly absorbed before they are capable of producing
energetic neutrinos.

\item There is also the possibility of searching for the annihilation products of standard, 
neutralino-like WIMPs using gamma ray telescopes. If the annihilation of WIMPs in the solar interior
creates large fluxes of neutrinos, the interaction of neutrinos in the upper layers of the solar atmosphere
will result in the production of hadrons ($\pi_0$, $K$,...) that decay with the significant yield of $\gamma$'s. The surviving 
$\gamma$ fraction which escapes may again be accessible to Fermi-LAT, and can provide alternative sensitivity to the 
WIMP-powered neutrino flux. While  ground-based neutrino telescopes are ideally suited to exploring 
$\nu_\mu$ neutrino fluxes, the gamma ray signature will contain information about other flavors, and 
be indirectly sensitive to the neutrino energy range where ground-based neutrino telescopes do not have 
any directional sensitivity. 

\item Another possibility that was not considered in this paper is WIMP annihilation to very light
quasi-stable  mediators, such as QCD axions. The conversion of axions into photons may occur in the magnetic field of the Sun, 
resulting in a $\gamma$-ray flux. This idea has some similarities with a recent proposal to 
observe the Sun's transparency to gamma ray sources that may contain an axionic component \cite{FRT}.

  \end{itemize}

In conclusion, the existence of a generic dark sector with metastable states mediating the
interactions between WIMP dark matter and the SM opens up new possibilities for indirect
detection signatures in the solar system. While it is possible to search for charged particles 
produced by mediator decays, it seems the hard gamma-ray flux produced by decays outside the
solar radius may be the most promising `smoking gun' signature due to its tight correlation with the solar center,
making it easily distinguishable from cosmic ray-induced backgrounds. As a final remark, we observe that
it may be profitable to explore other galactic or extra-galactic implications of this scenario in which
all stellar objects are effectively imbued with a high-energy gamma spectrum.

\subsection*{Acknowledgements}

The authors would like to thank M. Boezio, M. Casolino, D. Hanna and D. Hooper for helpful discussions 
and/or email correspondence. We also thank I. Yavin for informing us of his related work. B.B. acknowledges
support in part from the DOE under contract DE-FG02-96ER40969 during the 
Unusual Dark Matter workshop at the University of Oregon.
The work of A.R. and M.P. is supported in part by NSERC, Canada, and research at the Perimeter Institute
is supported in part by the Government of Canada through NSERC and by the Province of Ontario through MEDT.

\end{document}